\documentclass[twocolumn]{aastex631}

\usepackage{aas_macros}
\usepackage{soul}
\usepackage{amsmath}
\usepackage{ulem}

\begin{document}
\title{An experiment to observe GNSS signals with the Australian VGOS array}

\author[0000-0002-7563-9488]{Lucia McCallum}
\affiliation{University of Tasmania \\
School of Natural Sciences \\
Hobart, TAS 7005, Australia}

\author[0000-0003-2601-1413]{David Schunck}
\affiliation{University of Tasmania \\
School of Natural Sciences \\
Hobart, TAS 7005, Australia}

\author[0000-0002-0233-6937]{Jamie McCallum}
\affiliation{University of Tasmania \\
School of Natural Sciences \\
Hobart, TAS 7005, Australia}

\author[0000-0001-9525-7981]{Tiege McCarthy}
\affiliation{University of Tasmania \\
School of Natural Sciences \\
Hobart, TAS 7005, Australia}



\begin{abstract}

This paper introduces a new instrument enabling a novel combination of Earth measuring techniques: direct observations with the radio astronomical instruments to satellites of the global navigation satellite systems.
Inter-technique biases are a major error source in the terrestrial reference frame. Combining two major space-geodetic techniques, GNSS and VLBI, through observations to identical sensors has been considered infeasible due to their seemingly incompatible operating frequencies. 
The newly accessible L-band capability of the Australian VGOS telescopes is shown here, invalidating this prevailing opinion. A series of test observations demonstrates geodetic VLBI observations to GPS satellites for a continental-wide IVS telescope array, with the potential for observations at a critical scale. We anticipate immediate impact for the geodetic community, through first-ever inter-technique ties between VLBI and GNSS in the Australian region and via the opportunity for critical test observations towards the Genesis mission, geodesy's flagship project in the area of space ties set for launch in 2028.

\end{abstract}

\keywords{VLBI,  Astronomical Instrumentation}


\section{Introduction} \label{sec:intro}

Accurate timing, navigation and positioning information underpins modern society \citep{hiddenrisk2014}. It is provided through Global Navigation Satellite Systems (GNSS), such as the Global Positioning System (GPS). Satellite navigation itself relies on accurate orbit information. For this, one needs to disentangle the movement of the satellite in the Earth’s gravity field and the variable rotation of the Earth. Precise Earth orientation parameters are provided through the International Earth Rotation and Reference System Service (IERS), which combines measurements from the Global Geodetic Observing System (GGOS), such as those provided by GNSS or Very Long Baseline Interferometry (VLBI). In VLBI \citep[i.e.][]{nothnagel2017,sovers1998}, a global network of radio telescopes receives signals emitted from extragalactic radio sources, allowing for precise measurement of the distance between the receiving stations (baselines) as well as the orientation of the Earth with respect to a celestial reference frame \citep[ICRF;][]{charlot2020}. UT1-UTC, the difference between the actual rotation of the Earth and universal time maintained by a set of atomic clocks, is the parameter that is uniquely measured by the VLBI technique.

Geodesy is the science of measuring the geometric shape of the Earth, its orientation in space and its gravity field \citep{poutanen2020}. Its main product is the International Terrestrial Reference Frame \citep[ITRF2020;][]{altamimi2023}, the most accurate coordinate system of the Earth, underpinning all geosciences. The ITRF is a combined product of four different space-geodetic techniques, GNSS, VLBI, Satellite Laser Ranging (SLR) and Doppler Orbitography and Radiopositioning Integrated by Satellite (DORIS). Its accuracy (long-term origin) is estimated to be at the level of 5~mm \citep{altamimi2023}, which is a factor of five worse than today's needs \citep{ggos}. The restricting factor for further improvements of the ITRF is thought to be systematic errors of the individual techniques. The linking of the various techniques is crucial, which is usually done via local tie survey at co-location observatories, hosting components of multiple techniques \citep[e.g.][]{matsumoto2022}. In ITRF2020, tie discrepancies of up to several centimetres are present, thus limiting the overall accuracy of the coordinate frame.
This problem has long been recognised, searching for alternative methods to link the various techniques. SLR, GNSS and DORIS are all satellite techniques and people are working on combining them at the observation level   \citep[e.g.][]{thaller2011,flohrer2011,mannel2017,bury2021}. VLBI is different. Natural radio sources in use are extremely distant and emit over a wide range of frequencies, with signal strengths at a typical level of 1/1000 of the system noise. Geodetic VLBI was first used in the 1970s \citep{hinteregger1972}. The frequencies in use developed from globally available instruments and they are recognised and partly protected by the international telecommunications union (ITU). Currently, geodetic VLBI is done at S (2.2-2.4~GHz) and X (8.2-8.9~GHz) bands \citep{sovers1998, schuh2012} with the new generation VLBI Global Observing System \citep[VGOS;][]{niell2018} making use of the wideband frequency range between 3 and 14 GHz. In contrast, GNSS systems typically operate at L-band (1.2 and 1.6~GHz), emitting artificial signals (narrow-band), many orders of magnitude stronger than natural radio sources. Thus, intrinsically, geodetic VLBI and GNSS have incompatible observing requirements.

While there have been attempts in observing GNSS  satellites with VLBI \citep{tornatore2014, haas2014,haas2017,plank2017}, such investigations were always understood as pathfinders. VLBI telescopes and equipment are not designed to track GNSS satellites or receive and process their signals. When successful \citep[e.g.][]{plank2017}, the used telescopes were not the core instruments used for the TRF observations and standard processing chains revealed insufficiencies in handling the powerful narrow-band signals inherent to GNSS. A thorough combination between VLBI and GNSS can only occur if the same equipment and processes are used (telescopes, correlation and analysis software) as is used by the standard geodetic VLBI chain coordinated by the International VLBI Service for Geodesy and Astrometry \citep[IVS;][]{nothnagel2017}.   
An exception to this statement are efforts at Wettzell, where indeed L-band signals could be received using the standard S-band receiver \citep{haas2014}. A similar approach is followed for our work.

Accepting the fact that VLBI and GNSS cannot simply be connected via direct observation of the L-band signal with the IVS network, alternative systems were suggested: \citet{rothacher2009} defined the concept of a space tie, combining receivers/transmitters of all four techniques on a single satellite platform. There were mission proposals submitted to space agencies, such as GRASP \citep{barsever2009} and E-GRASP/Eratosthenes \citep{biancale2017} before ESA selected the 
Genesis mission \citep{delva2023}. It is also worth mentioning NASA’s GRITSS project, which aims to link the VLBI and GNSS systems by using a dedicated satellite in a heterodyning approach.

The VLBI team of the University of Tasmania, for the first time, demonstrate that a geodetic VLBI array can observe the GNSS constellation in L-band and derive geodetic observations, in a (close to) production mode. Geodetic VLBI in Australia is operated by the University of Tasmania, contracted through Geoscience Australia. There are three VGOS telescopes on the continent, in Hobart (Tasmania), Katherine (Northern Territory), and Yarragadee (Western Australia), forming the AuScope VLBI array.  In September/October 2024 we observed a set of GPS satellites over several hours, processed the observations through the standard VLBI processing chain deriving geodetic baseline delays. These successful observations are a surprise to the community, which has had the belief that modern VGOS systems are unable to receive GNSS signals.

The consequences of this newly developed technique are twofold: A direct link between VLBI and GNSS will be possible, initially over Australia, and globally if this technique can be transferred to other VGOS telescopes worldwide; presenting a new opportunity with significant implications for an improved TRF. Secondly, this proof-of-concept is just the important piece missing in order to make the VLBI part of the Genesis mission a success: after the mission being granted it has become obvious that the VLBI observations to Genesis are completely novel and in order to be successful, a multitude of technological challenges must be overcome. This becomes achievable with  actual field tests. 

 \subsection{State of the art}
 VLBI observations to spacecrafts are routinely included in (outer) space missions, leveraging on the high precision measurements for orbit determination \citep{duev2012,hanada2008,lebreton2005}. When it comes to geodetic VLBI though, incorporation of novel observations is not straightforward. Observing the Chinese Chang'e3 lunar lander with the IVS network \citep{klopotek2019} and VLBI observations from the AuScope array to the APOD mission \citep{sun2018, hellerschmied2018}, revealed some of these issues. One of the conclusions from these studies was that optimisation of the technique will require more observations.
 \citet{boehm2024} nicely summarise current work towards VLBI transmitters on satellites. Multiple groups have studied this technique using simulations \citep{plank2014, anderson2018, klopotek2020, wolf2023, pollet2023, schunck2024a, schunck2024b} with studies focusing on the technical realisation by \citet{jaradat2021}, \citet{plank2017}, and \citet{schunckIAG}. A slightly different observing concept using new hardware is discussed by \citet{skeens2023}. 
 
The current knowledge gap is in the full implementation of these observations into routine operations and procedures, resulting from a lack of opportunity to develop the technique due to poor frequency compatibility between existing satellites and the majority of radio telescopes. Before VLBI observations to GNSS satellites can be scientifically useful, proof of quality measurements in terms of a geodetic delay without suspicious biases needs to be achieved. This includes multi-frequency delays in order to correct for ionospheric effects and suitable strategies on how to deal with the artificial signal in the fringe fitting procedures as well as the various polarisation products arising from the use of dual-linear feeds.

\section{Demonstrator experiments} \label{sec:demonstrator}

\subsection{AuScope VLBI Array}
\begin{figure}[t]
    \centering
    \includegraphics[width=\linewidth]{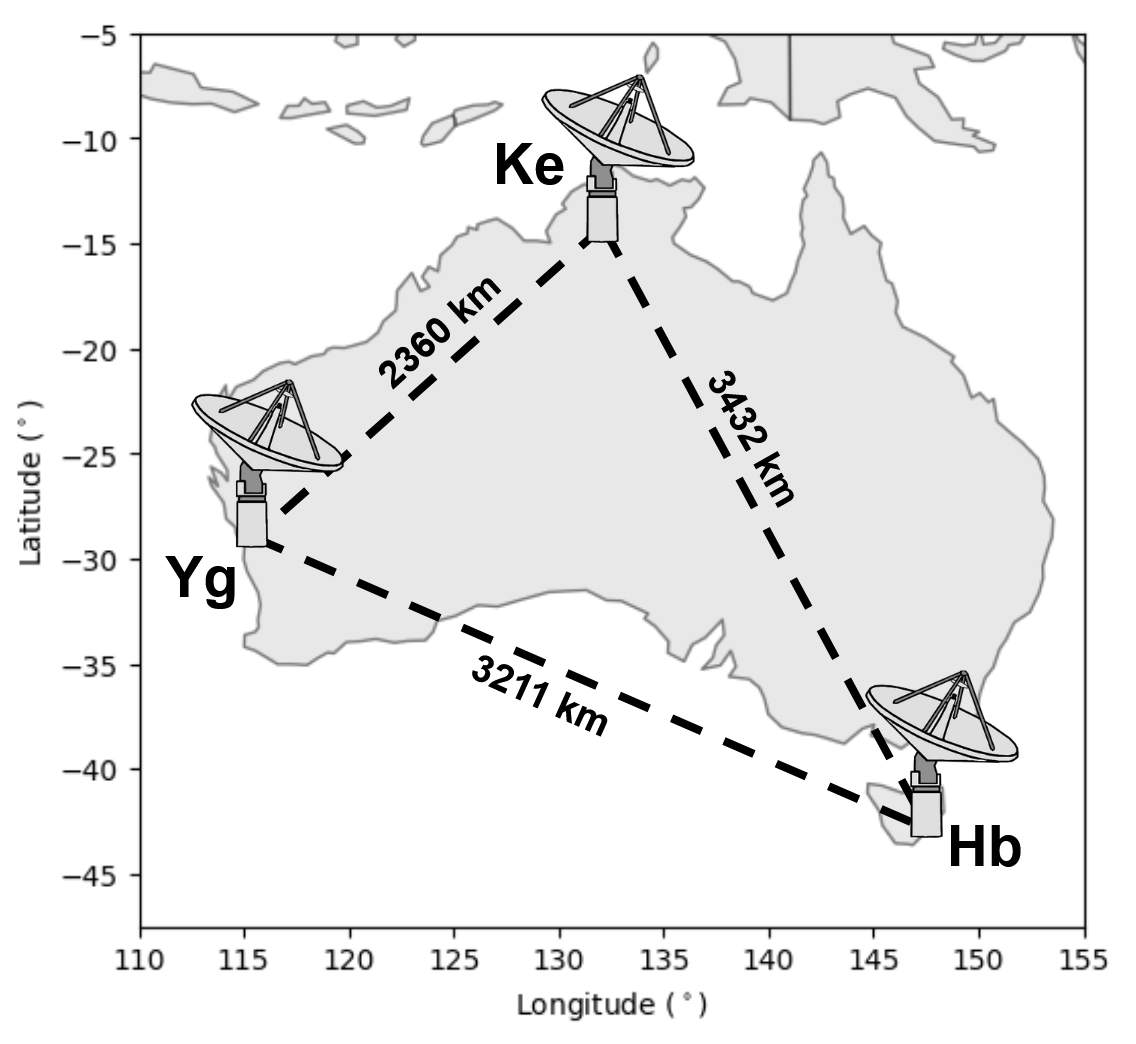}
    \caption{AuScope VLBI array. Stations are located across the Australian continent, in Hobart (Hb), Katherine (Ke), and Yarragadee (Yg).}
    \label{fig:array}
\end{figure}
 The AuScope VLBI array \citep{lovell2013} is comprised of three 12m telescopes across the Australian continent (Fig \ref{fig:array}).
 The telescopes are owned and operated by the University of Tasmania, with operational IVS observing funded through Geoscience Australia. Initially operating as an S/X legacy array \citep{plank2015, plankAUSTRAL} the telescopes have now been upgraded to VGOS receivers and signal chains. The implementation of these VGOS upgrades is different compared to other similar stations, which allows for increased flexibility in the observing modes \citep[e.g. demonstrated in][]{mccallum2022}. For details about the technical specifications of the telescopes and receiver design the reader is referred to \citet{lovell2013} and \citet{mccallum2022}.

 \subsection{Observations}
 For the test observations, three satellites of the GPS system (block IIF) were observed. The AuScope array has QRFH VGOS feed horns \citep{akgiray2013}, with a nominal operating frequency range between 2-14~GHz. The key discovery for this work was that despite the nominal cut-off, GPS signals in L-band were detected with VLBI. This is due to the fact that GPS signals are tremendously strong. After the feed horn and standard VGOS cryogenic low noise amplifiers (LNA) the signal is split. The satellite signal passes through a second-stage LNA and filter stage (500-1650~MHz) and travels via coaxial cable without down-conversion to the control room. The observed frequencies lie out of range for the normal calibration techniques in place at the VGOS telescopes. The approximate sensitivity of our receiving system is on the order of several 100.000~Jy in source equivalent flux density (SEFD).
 
 For these tests, we typically used blocks of a few hours on either 2 or 3 telescopes. The preparation of these experiments required is minimal (1-2 hrs). Observing itself is done remotely, with the current bottleneck being the data transport to the correlator in Hobart; while Yarragadee data can be e-transferred at good speeds, the data from Katherine needs to be physically shipped until the high speed data connection will become available during 2025. For future full scale sessions (24-hrs) we would expect turnaround times of about 2-3 weeks, which is common for IVS observations.

 VGOS feed horns are linearly polarised. This means, that we record the circularly polarised signal from the satellite in two polarisations at each station. During correlation, the two data streams from each station are cross-correlated, leading to four correlation products, i.e. XX, XY, YX, and YY. Similar to the approach of \citet{plank2017}, all subsequent investigations are using the XX product only. A thorough combination and calibration for the full polarisation product could be subject of future research. Using all data streams will certainly increase the sensitivity of the system, which is not a limiting factor at the moment. A more serious issue would be if delay offsets were found between the various products, which is yet to be investigated in detail.
 
\begin{figure}
    \centering
    \includegraphics[width=\linewidth]{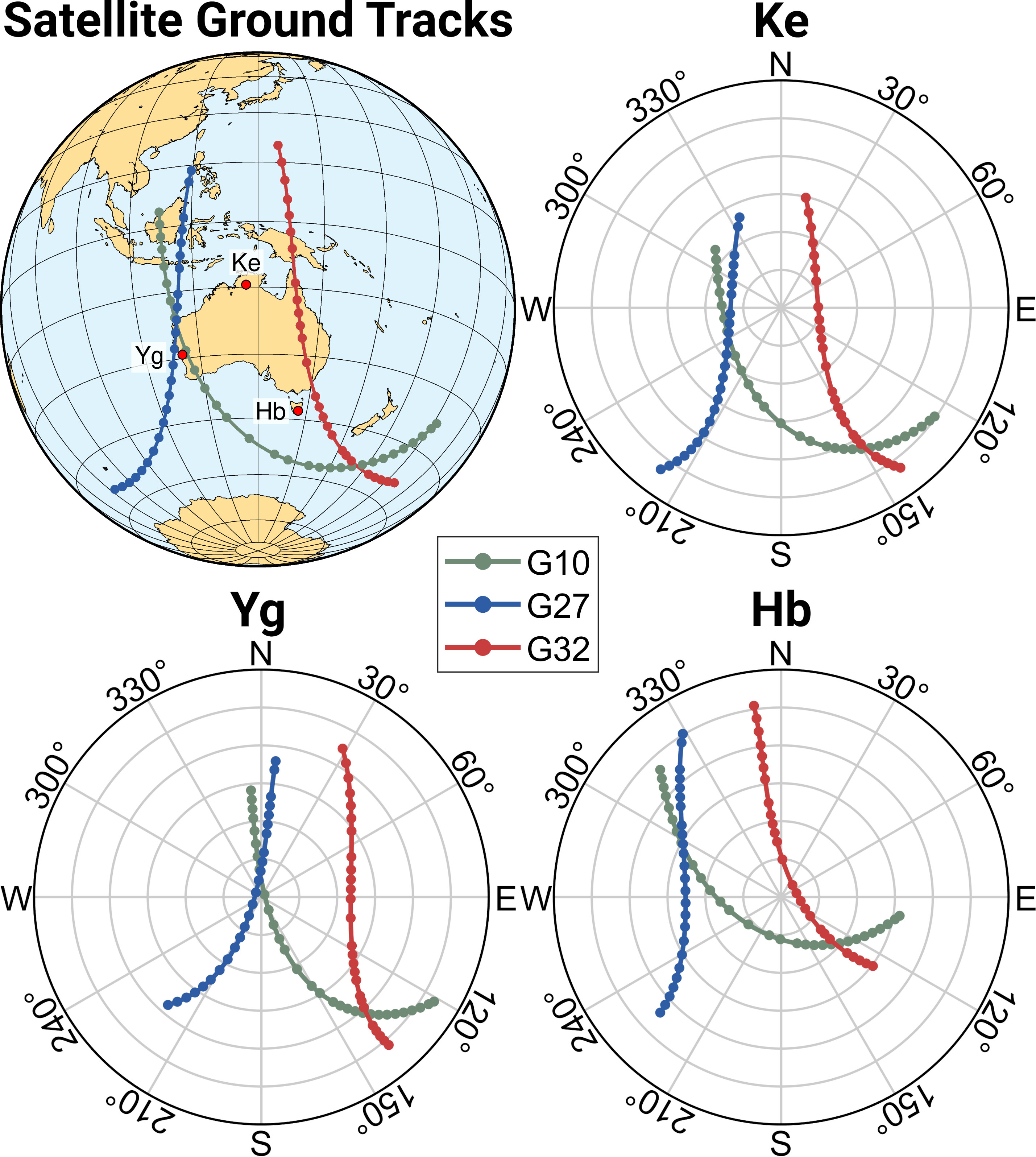}
    \caption{Geometry of experiment ASO304. The satellite ground tracks are shown for the three observed satellites along with topocentric view angles at the observing stations.}
    \label{fig:skyplot}
\end{figure}

The main experiment investigated below is ASO304, observed on October 30, 2024. Figure \ref{fig:skyplot} shows the geometry of this experiment. The observations are split into individual scans, integrating over the scan duration of 20~seconds. Three GPS block IIF satellites (G10, G27, G32) were observed, with a minimal repeat time of the same satellite of 7~minutes.

\begin{figure*}
    \centering
    \includegraphics[width=.8\textwidth]{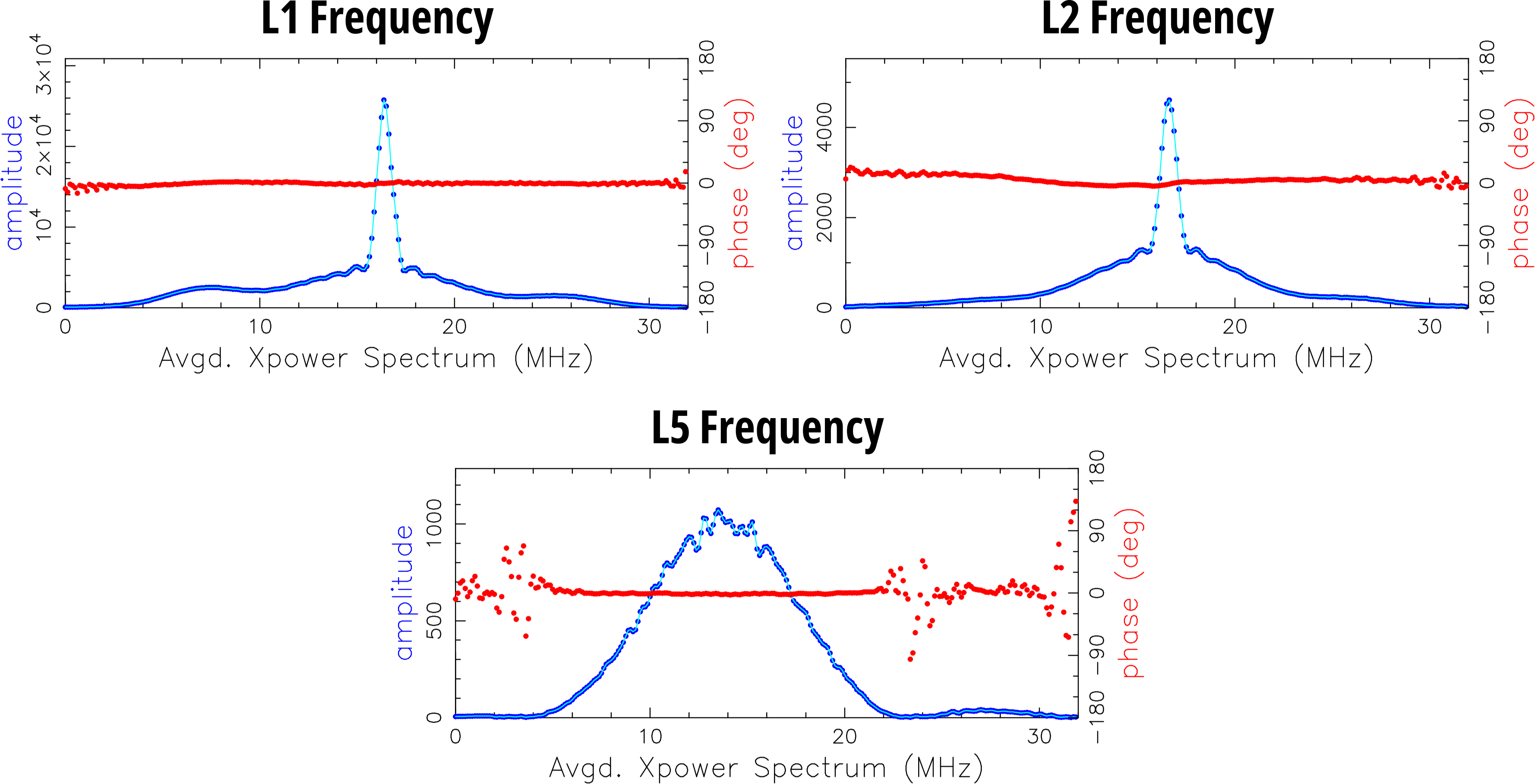}
    \caption{Fringe plot for experiment ASO304 using the XX cross polarisation product for baseline Katherine-Yarragadee. Shown are the residual phase and averaged amplitude for the three recorded frequencies in a 20 second integration. The frequency axis is an offset to the zoom-band edge, corresponding to sky frequencies of 1163.0, 1211.0, and 1559.0~MHz for L5, L2 and L1 respectively.}
    \label{fig:fringe}
\end{figure*}
 
\subsection{Key developments}

In the presented observations, a number of novelties have been applied:

\subsubsection{Multi-frequency:}
Despite using out-of-frequency receivers, GPS signals on the three main frequencies, L1, L2 and L5, could be observed. These out-of-band signals are strongly attenuated by the receivers, which proves beneficial for the subsequent VLBI processing. Indeed, these observations do not cause significant variations in the total power received at the first-stage amplifiers or at various stages in the receiving system. This is in stark contrast to previous observations by \citet{plank2017}, where strong signals partially caused saturation or at least problems with the automated gain control systems in the digitization stage. We thus expect better VLBI performance for the new observations. 
Successful observations were also done to satellites of the Galileo (E5a, E5b, E6 and E1) and Beidou (B2a, B2b, B3 and B1) system at sky frequencies of 1176.45, 1207.14, 1268.52, and 1575.42~MHz respectively. In general, we expect compatibility with all GNSS systems.

\subsubsection{VGOS telescopes and recorders:}
Previous VLBI observations to satellites of the GNSS were done using astronomical radio telescopes or at least different (L-band) receivers or other modifications. This time the standard VGOS equipment is used. In our case, this means a VGOS receiver, sampling using the DBBC-3\footnote{digital base band converter, \url{https://www.hat-lab.cloud/dbbc3-2/}} at sky frequencies and recording with a FlexBuff \citep{Lindqvist2014} machine. This, in principle, would allow combination with actual VGOS observations to natural radio sources within the same experiment. The implementation of this mode change is something that requires further study and testing but seems readily achievable. The same challenges apply to observations of the Genesis satellite.

\subsubsection{Geodetic observing and processing pipeline:}
In contrast to test observations, for bulk data production one has to make sure that all observing and processing goes through standard geodetic/IVS procedures and software. With these tests, we have made new steps towards full incorporation into the geodetic pipeline. An observing plan can be created using state-of-the-art scheduling software, i.e. VieSched++ \citep{schartner2019} which allows for the option of satellite targets. As the \textit{.skd} format (which is typically used for observing) does not support satellite targets yet, we create so-called snap-files (\textit{.snp}). These can be loaded into the antenna field systems (NASA field system\footnote{\url{https://github.com/nvi-inc/fs}}) that is used to coordinate the observations including antenna steering and recording. 
Since the AuScope telescopes are not set up for continuous satellite tracking via the field system, we use the \textit{satellite} command (in the field system), implementing step-wise tracking. An update interval of 5 seconds was found to be sufficient for the GNSS satellites to stay within the main beam of the telescopes. For the conducted experiments, a maximal angular change of $0.046^{\circ}$ per 5~second update interval was found. This is one magnitude lower than the half power beam width of $0.91^{\circ}$ calculated for the 12~m telescopes at L1.

For correlation using DiFX \citep{deller2007,deller2011}, the a priori delay model is created through \textit{difxcalc} via externally created SPICE Kernel files \citep{acton1996, acton2018}. While the correlation process delivers visibilities, the actual VLBI delay is formed in a process typically referred to as fringe-fitting using the \textit{fourfit} program as part of the HOPS \citep{MIThops} pipeline. Here some minor tweaks had to be made to pass validation checks when using 8-bit sampled data.

\subsubsection{8-bit recording:}
Digitization at an adequate bit-depth is an essential step in the VLBI technique. A higher recording rate corresponds to higher signal to noise ratios which in turn increase the precision in the final measurement. In geodetic VLBI, typically 2-bit sampling is used, sufficient for the very weak, noise-like natural radio signals. As shown in \citet{plank2017}, the characteristics of the GNSS signals with a very pronounced main peak and side lobes causes issues in the fringe fitting stage leading to variations in the estimated geodetic delay. They recommend that adopting a higher bit rate could improve the dynamic range of the recorded data, reducing the effects of compression from the strong transmitter tones.
Rather novel for geodetic VLBI, most of our test observations were made with 8-bit recording. The three primary GPS signals were observed across two 128~MHz channels (L1 in the first, L2 and L5 in the second). 
In Figure \ref{fig:fringe} the so-called fringe-plots are shown for the three observed frequencies. Zoom-bands of 32~MHz were used to concentrate on the signal-carrying portion of the spectrum. In the amplitudes, the signal structure characteristic for L1, L2 and L5 is clearly visible.
The flat phase residuals indicate a good fit to the measured delay. Observations were also successfully done using the standard 2-bit recording. A thorough investigation into the benefits of the more data-intensive 8-bit mode compared to the 2-bit mode is pending.

It shall further be noted, that some observations did not lead to a reliable fringe detection and delay measurement. Those were mostly related to satellites of the block III type, with a different signal modulation. Most likely, the characteristic double-peak related to the signal modulation is the reason for unsuccessful delay derivation, though it is also possible that the way the fringe fitting is applied or the lack of proper polarisation handling are causing the issues here. It does emphasize that further research is necessary to account for a variety of GNSS signals.

 \section{Results}
 The result of these experiments is the finding that the AuScope VLBI array is an instrument that can perform VLBI observation to GNSS satellites. Such observations can be done without significant changes to the VGOS setup, thus at short notice, and at scale.  \\
 The target observables in a VLBI experiment are baseline delays, representing the geometric travel time (total delays) between the signal arriving at telescope one and telescope two. These time delays are the result of the correlation and fringe-fitting process and are the input data for subsequent geodetic analysis \citep[e.g.][]{schuh2013}. 
 
\begin{figure*}[t]
    \centering
    \includegraphics[width=.8\textwidth]{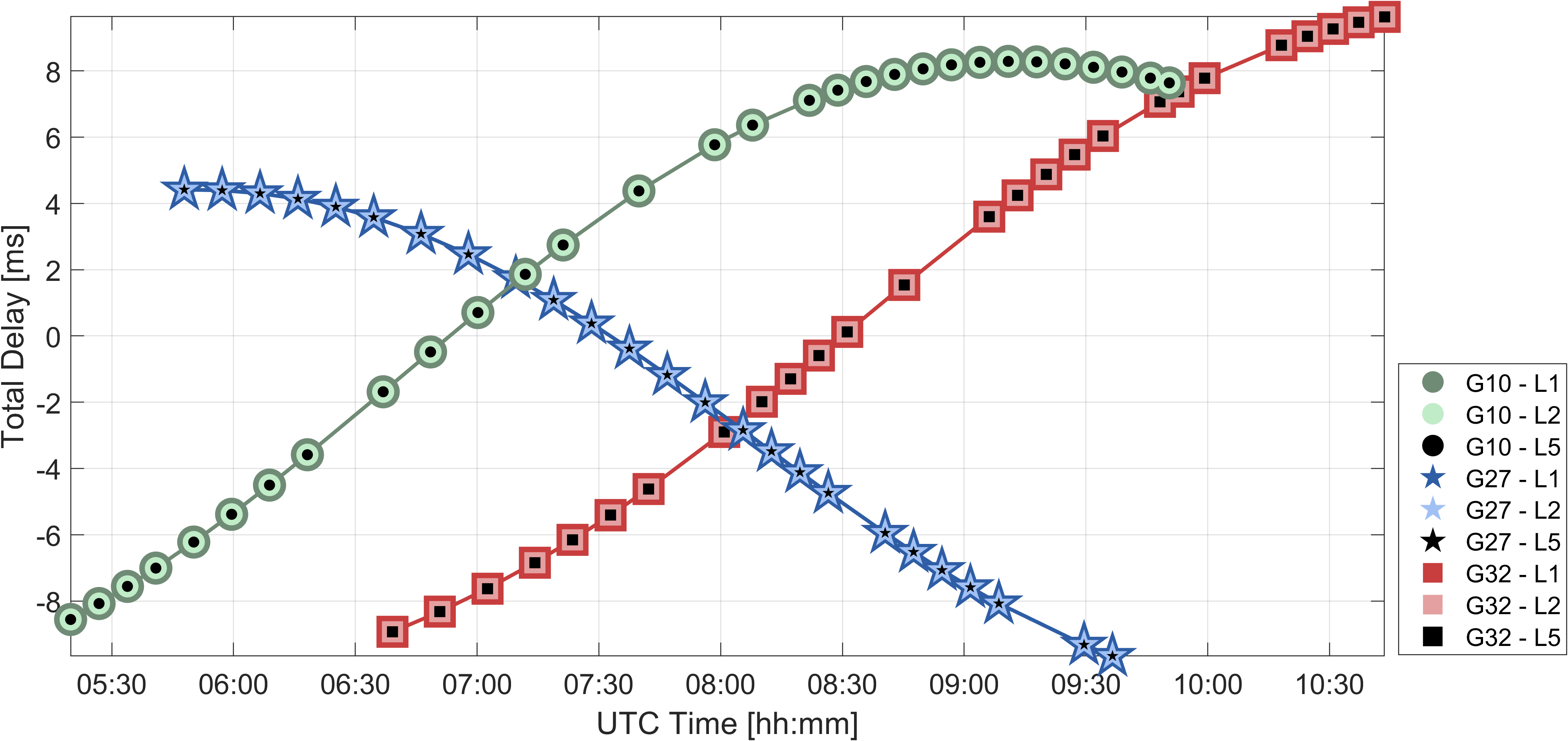}
    \caption{Total delays of experiment ASO304 for the baseline Hobart-Katherine. Each data point represents one measurement from a 20~sec scan. Three satellites of the GPS were observed (G10, G27, G32), with delays measured in L1, L2 and L5.}
    \label{fig:totaldelay}
\end{figure*}

\begin{figure*}[t]
    \centering
    \includegraphics[width=1.0\textwidth]{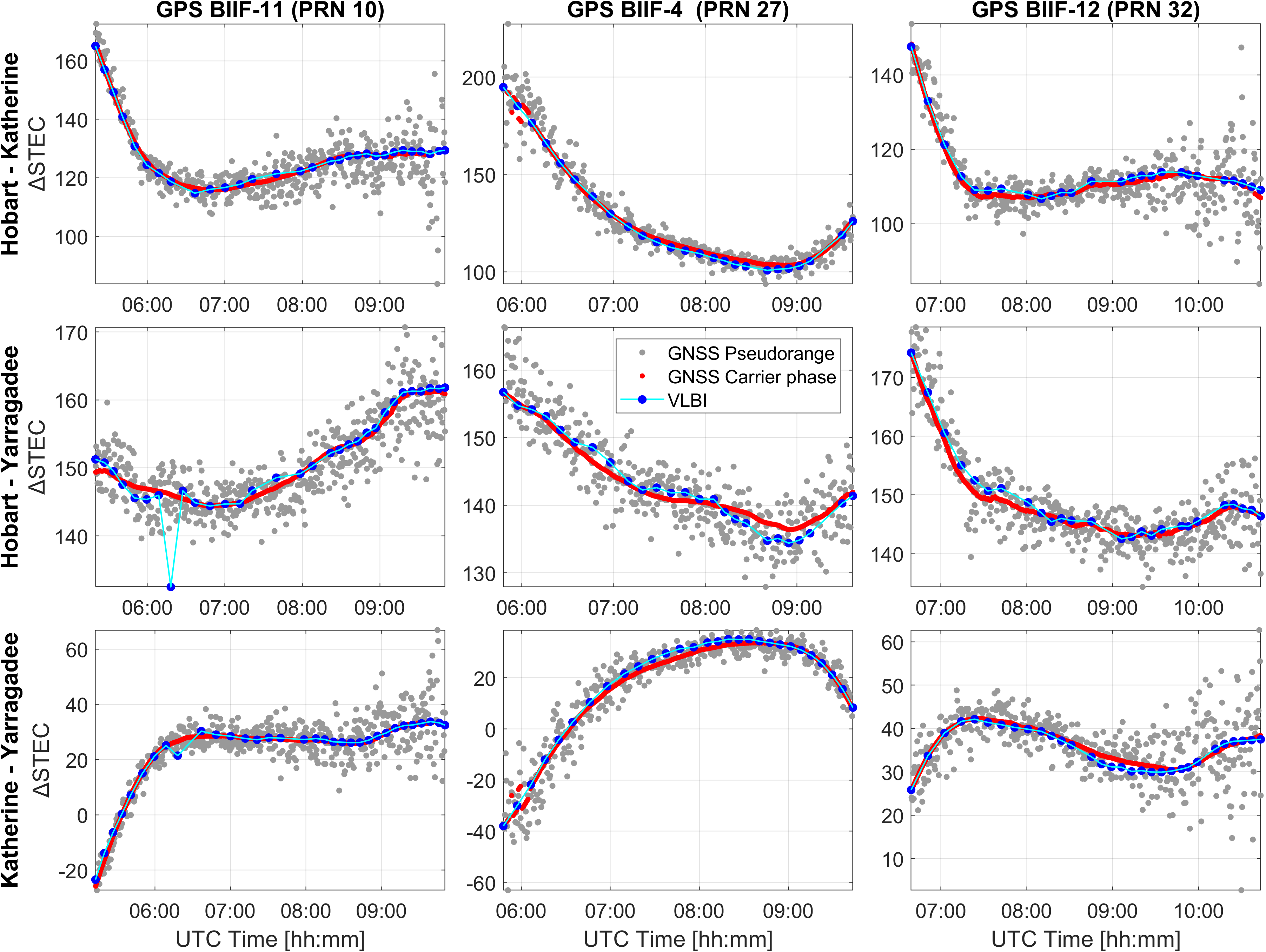}
    \caption{Differential slant total electron content ($\Delta$STEC) measured in TEC units for ASO304 observing three satellites. The results are shown for three baselines, comparing the values derived using the linear combination of L1 and L2 from the VLBI delays (\textit{blue}), as well as the GPS results calculated from the pseudoranges (\textit{gray}) and the carrier phase (\textit{red}). One constant offset was estimated and removed for each baseline and satellite for the VLBI data. The full mathematical formulations are given in the appendix.}
    \label{fig:iono}
\end{figure*}

In figure \ref{fig:totaldelay} we show the total delays of experiment ASO304, exemplarily for one of the three baselines. Three satellites were observed alternately over a total duration of 6 hours, with 20~seconds scan duration on the satellites. One clearly sees a smooth variation of the delays over time, due to the satellite's orbital movement. 

Having observations in different frequencies (L1, L2, L5) allows to measure the dispersive time delay caused by the ionosphere. By forming the ionosphere-free linear combinations (see Appendix), we can measure the ionospheric delay for each satellite. These values are then compared to those derived directly from GPS observations, as shown in Figure \ref{fig:iono}. We find good agreement between the VLBI and pure GPS results.

A first attempt for analysis of the data was done using the VieVS \citep{boehm2018} environment, which has a satellite model implemented. Comparing the observed and computed delays we find residuals at the level of a few nanoseconds. The residuals are at a similar level as reported in \citet{plank2017}, though this time there are observations on three baselines compared to their single baseline observations. The residuals are one order of magnitude better to the correlator model using \textit{difxcalc}. 
Once more observational data is available, we expect the analysis to provide meaningful geodetic results. The current residuals are thought to mainly be due to uncalibrated clock errors, the troposphere as well as refinements in the delay model that is implemented in the software. In addition, as pointed out throughout this paper, there is yet more work to be done to further understand the VLBI processing (i.e. correlation and fringe fitting process) and delay generation of this unusual signal.

\section{Discussion}

This new instrument with L-band capability has two direct consequences for the geodesy community. Generating a critical dataset of VLBI observations to GNSS satellites will allow, for the first time, scientific use of such data. Direct comparison between VLBI-results, GNSS-results, and local ties in the test region (i.e. Australia) will allow for a search of technique-specific biases. With regard to realising a space-tie, there are actions in the community to develop new methods and software for combined orbit-determination, also including, for the first time, VLBI observations. Availability of a real dataset is important to accelerate these efforts. The second major application of this new instrument is development for the Genesis mission. It is clear that there are many open questions for the VLBI component. With the newly available capability for extensive test observations, hopefully many of those can be resolved before start of the mission in 2028.  

 Finally, we would like to stress that there is yet more development necessary to make VLBI observations to GNSS satellites precise and accurate enough. And we are excited to report that there now is an instrument available which allows these developments to happen. 

\section*{Acknowledgments}
The AuScope VLBI project is managed by the University of Tasmania, contracted through Geoscience Australia.
This work was supported by the Australian Research Council (DE180100245). 
This research has made use of NASA’s Astrophysics Data System Abstract Service. 

\appendix
\section{Calculation of Differential Slant TEC}

\subsection{VLBI}
In the following the estimation of the differential slant TEC from geodetic VLBI observations is described. We follow the formulation given in \citet{Alizadeh2013}. The observed group delays of the L1 and L2 frequencies can be described as
\begin{eqnarray}
	\tau_{L1} &= \tau_{if} + \dfrac{\alpha}{f^2_{L1}},& \hspace{2em} \text{and}\label{eq:A1}\\
	\tau_{L2} &= \tau_{if} + \dfrac{\alpha}{f^2_{L2}},& \label{eq:A2}
\end{eqnarray}

where $\tau_{if}$ is the ionospheric free delay and $\alpha$ is the contribution from the ionosphere, scaled by the corresponding effective ionosphere frequencies $f_{L1}$ and $f_{L2}$. The first-order approximation of the constituent $\alpha$ is given by
\begin{equation}
	\alpha = \dfrac{40.31}{c} \left( \int N_e ds_1 - \int N_e ds_2 \right) = \dfrac{40.31}{c} (STEC_1 - STEC2), \label{eq:A3}
\end{equation}

where $s_1$ and $s_2$ are the paths of wave propagation from the source to the first and second receiving antenna, $STEC_i$ is the integrated electron density $Ne$ along the ray paths $s_1$ and $s_2$, and $c$ is the speed of light used for conversion to time delay. Using \ref{eq:A1} and \ref{eq:A2}, the ionospheric free delay observable $\tau_{if}$ can be eliminated and the unknown parameter $\alpha$ can be obtained by a linear combination
\begin{equation}
	\alpha = \dfrac{f^2_{L1} f^2_{L2}}{f^2_{L2}-f^2_{L1}} (\tau_{L1} - \tau_{L2}). \label{eq:A4}
\end{equation}

By combining \ref{eq:A3} and \ref{eq:A4}, the differential slant TEC is given by
\begin{equation}
	\Delta STEC = STEC_1 - STEC2 = \dfrac{c}{40.31} \dfrac{f^2_{L1} f^2_{L2}}{f^2_{L2}-f^2_{L1}} (\tau_{L1} - \tau_{L2}). \label{eq:A5}
\end{equation}

In contrast to \ref{eq:A1} and \ref{eq:A2}, real observations include an additional delay introduced by instrumental delays in the different bands, changing the delay observable to
\begin{eqnarray}
	\tau^{\prime}_{L1} = \tau_{if} + \dfrac{\alpha}{f^2_{L1}} + \tau_{inst,L1},& \hspace{2em} \text{and}\label{eq:A6}\\
	\tau^{\prime}_{L2} = \tau_{if} + \dfrac{\alpha}{f^2_{L2}} + \tau_{inst,L2}.& \label{eq:A7}
\end{eqnarray}

If the instrumental delays do not change over time, they cause a constant bias in the determination of the differential slant TEC.

\subsection{GNSS}
In the following the estimation of the differential slant TEC from GNSS observations is described, following \citet{lin1997}. At one GNSS station, the difference of the pseudo-range observations in the L1 and L2 frequencies are directly proportional to the slant TEC between the satellite and the antenna. The slant TEC at station $i$ with pseudo-range measurements is given by
\begin{equation}
	STEC_{P,i} = \dfrac{1}{40.31} \dfrac{f^2_{L1} f^2_{L2}}{f^2_{L2}-f^2_{L1}} (P_2 - P_1), \label{eq:A8}
\end{equation}

where $P_1$ and $P_2$ are the pseudo-range observations at the frequencies $f_{L1}$ and $f_{L2}$, respectively. To compute the differential slant TEC, the slant TEC values are determined at two stations and then subtracted from one another. While the difference of the pseudo-ranges has all frequency-independent effects removed, such as satellite motion, tropospheric effects and clock errors, other errors, such as multipath as well as receiver and satellite L1/L2 differential delays, cause errors in the estimation. In contrast to pseudo-range measurements, carrier phase observations have less measurement noise. However, they contain integer ambiguities which cause an unknown bias. The estimation of the slant TEC at station $i$ with carrier phase measurements is given by
\begin{equation}
	STEC_{\Phi,i} = \dfrac{1}{40.31} \dfrac{f^2_{L1} f^2_{L2}}{f^2_{L2}-f^2_{L1}} (\Phi_2 - \Phi_1) + D, \label{eq:A9}
\end{equation}

where $\Phi_1$ and $\Phi_2$ are the carrier phase measurements at the frequencies $f_{L1}$ and $f_{L2}$. The unknown bias $D$ is described by
\begin{equation}
	D = - \dfrac{1}{40.31} \dfrac{f^2_{L1} f^2_{L2}}{f^2_{L2}-f^2_{L1}} (\lambda_2 N_2 - \lambda_1 N_1), \label{eq:A10}
\end{equation}

where $\lambda_1$ and $\lambda_2$ are the carrier wavelengths at L1 and L2 respectively; $N_1$ and $N_2$ are the corresponding integer ambiguity parameters. Without determining $D$, \ref{eq:A9} provides only a relative measurement of the TEC. Accurate absolute values of TEC are obtained by combining pseudo-range and carrier phase measurements.

\bibliography{references}{}
\bibliographystyle{aasjournal}



\end{document}